\documentclass[12pt]{article}%
\usepackage{amsmath}
\usepackage{amsfonts}
\usepackage{amssymb}
\usepackage{graphicx}%
\providecommand{\U}[1]{\protect\rule{.1in}{.1in}}
\newtheorem{theorem}{Theorem}[section]

\newtheorem{remark}[theorem]{Remark}

\numberwithin{equation}{section}

\begin{document}

\title{Spectral parameter power series for Zakharov-Shabat direct and inverse
scattering problems}
\author{Vladislav V. Kravchenko\\{\small Departamento de Matem\'{a}ticas, Cinvestav, Unidad Quer\'{e}taro, }\\{\small Libramiento Norponiente \#2000, Fracc. Real de Juriquilla,
Quer\'{e}taro, Qro., 76230 MEXICO.}\\{\small e-mail: vkravchenko@math.cinvestav.edu.mx}}
\maketitle

\begin{abstract}
We study the direct and inverse scattering problems for the Zakharov-Shabat
system. Representations for the Jost solutions are obtained in the form of the
power series in terms of a transformed spectral parameter. In terms of that
parameter, the Jost solutions are convergent power series in the unit disk.
The coefficients of the series are computed following a simple recurrent
integration procedure. This essentially reduces the solution of the direct
scattering problem to the computation of the coefficients and location of
zeros of an analytic function inside of the unit disk.

The solution of the inverse scattering problem reduces to the solution of a
system of linear algebraic equations for the power series coefficients, while
the potential is recovered from the first coefficient.

The overall approach leads to a simple and efficient method for the numerical
solution of both direct and inverse scattering problems, which is illustrated
by numerical examples.\ 

\end{abstract}

\section{Introduction}

The system introduced by Zakharov and Shabat in \cite{ZS} for developing the
inverse scattering transform method for the nonlinear Schr\"{o}dinger
equation, has multiple applications. It is used for solving nonlinear
evolution equations (the nonlinear Schr\"{o}dinger, sine-Gordon and modified
Korteweg - de Vries equations) \cite{Ablowitz Segur}, \cite{Lamb}, and also
for modelling other physical phenomena (see, e.g., \cite{Sacks}, \cite{Tang et
al}). Efficient methods for solving both direct and inverse scattering
problems for the Zakharov-Shabat system with potentials decaying at infinity
are subject of numerous publications (see, e.g., \cite{Arico et al 2011},
\cite{Delitsyn 2022}, \cite{Fermo et al 2016}, \cite{Frangos et al 1991},
\cite{Frumin et al 2015}, \cite{Gorbenko et al}, \cite{Le et al 2014},
\cite{Mullyadzhanov et al 2021}, \cite{Xiao et al 2002} and the recent review
\cite{Medvedev et al 2023}). 

In this study, a completely different approach to the solution of direct and
inverse Zakharov-Shabat scattering problems is proposed and developed. This is
based on the following results. The Jost solutions of the Zakharov-Shabat
system admit power series expansions in terms of the parameter $z$, which is
related to the spectral parameter $\rho$ from the Zakharov-Shabat system (see
system (\ref{ZS1}), (\ref{ZS2}) below) by an elementary M\"{o}bius map
$z=\frac{\frac{1}{2}+i\rho}{\frac{1}{2}-i\rho}$. This transforms the upper
half-plane $\operatorname{Im}\rho\geq0$ onto the unit disk $\left\vert
z\right\vert \leq1$. For a given potential from the Zakharov-Shabat system,
the coefficients of the power series are computed following a simple recurrent
integration procedure. So, the approximate solution of the direct scattering
problem reduces to dealing with polynomials of the variable $z$. In
particular, the location of the eigenvalues reduces to finding roots of a
polynomial inside of the unit disk. 

Since $z$ in a sense is a spectral parameter as well, we call the power series
expansions of the Jost solutions as the spectral parameter power series
(SPPS). 

The inverse scattering problem reduces to a system of linear algebraic
equations for the SPPS coefficients, and moreover, the potential is recovered
from the very first SPPS coefficient. 

The SPPS representations for the Jost solutions arise from their integral
representations, as a result of a Fourier-Laguerre series expansion of the
integral kernels. This idea was proposed for the first time in
\cite{Kr2019MMAS InverseAxis} for application to the Schr\"{o}dinger equation
and was developed further in subsequent publications \cite{DKK2019MMAS},
\cite{DKK Jost}, \cite{GKT2023MMAS}, \cite{KarapetyantsKravchenkoBook},
\cite{KrBook2020}. Here, an analogous approach is applied to the
Zakharov-Shabat scattering problems. 

It is worth mentioning that the term SPPS was coined in \cite{KrPorter2010} in
relation to Sturm-Liouville problems on finite intervals. Here, we use it to
refer to the power series in terms of the mapped spectral parameter. 

The use of series representations in numerical analysis invokes always the
question as to how many terms should be used. We propose straightforward
criteria for the choice of this parameter both when solving the direct and
inverse problem. The criteria are based on some particular features of the
SPPS coefficients.

The overall approach to solving both direct and inverse Zakharov-Shabat
scattering problems lends itself to efficient numerical algorithms which are
tested on several numerical examples. 

In Section 2 we introduce the necessary notations and formulate the problems.
In Section 3 we introduce the SPPS representations for the Jost solutions and
prove their main properties. Sections 4 and 5 are dedicated to solving the
direct and inverse scattering problems, respectively. In Section 6 we discuss
the numerical implementation of the approach and illustrate it with several
numerical examples. Finally, Section 7 contains some concluding remarks.

\section{Statement of direct and inverse problem}

Let $q(x)$ be a real valued, piecewise continuously differentiable function
defined on $\left(  -\infty,\infty\right)  $. Consider the Zakharov-Shabat
system
\begin{equation}
\frac{dn_{1}(x)}{dx}+i\rho n_{1}(x)=q(x)n_{2}(x),\label{ZS1}%
\end{equation}%
\begin{equation}
\frac{dn_{2}(x)}{dx}-i\rho n_{2}(x)=-q(x)n_{1}(x)\label{ZS2}%
\end{equation}
where $\rho\in\overline{\mathbb{C}^{+}}:=\left\{  \rho\in\mathbb{C}%
\mid\operatorname{Im}\rho\geq0\right\}  $ is a spectral parameter. Although
the presented approach can be extended onto the Zakharov-Shabat system with a
complex potential, in the present work we consider the case of the real valued
one. This is because interesting relations between the SPPS coefficients arise
(subsection \ref{subsect SPPS for ZS}), leading to certain simplifications. 

For any two solutions of (\ref{ZS1}), (\ref{ZS2})
\[
\theta(x)=\left(
\begin{tabular}
[c]{c}%
$\theta_{1}(x)$\\
$\theta_{2}(x)$%
\end{tabular}
\ \right)  \quad\text{and}\quad\omega(x)=\left(
\begin{tabular}
[c]{c}%
$\omega_{1}(x)$\\
$\omega_{2}(x)$%
\end{tabular}
\ \right)  ,
\]
for the same value of the parameter $\rho$, the expression $W\left[
\theta;\omega\right]  :=\theta_{1}\omega_{2}-\theta_{2}\omega_{1}$ is
constant, $\frac{dW}{dx}=0$, and the nonvanishing of $W\left[  \theta
;\omega\right]  $ guarantees the linear independence of $\theta$ and $\omega$.

It is convenient to consider solutions of (\ref{ZS1}), (\ref{ZS2}) as
functions of $\rho$ as well, so we will write
\[
n=n(\rho,x)=\left(
\begin{tabular}
[c]{c}%
$n_{1}(\rho,x)$\\
$n_{2}(\rho,x)$%
\end{tabular}
\right)  .
\]

Denote
\[
\widetilde{n}(\rho,x):=\left(
\begin{tabular}
[c]{c}%
$n_{2}(-\rho,x)$\\
$-n_{1}(-\rho,x)$%
\end{tabular}
\right)  .
\]
It is easy to see that $n(\rho,x)$ and $\widetilde{n}(\rho,x)$ are solutions
of (\ref{ZS1}), (\ref{ZS2}) simultaneously.

We will suppose that $q(x)$ decays at $\pm\infty$ sufficiently fast, so that
there exist the unique, so-called Jost solutions
\[
\varphi(\rho,x)=\left(
\begin{tabular}
[c]{c}%
$\varphi_{1}(\rho,x)$\\
$\varphi_{2}(\rho,x)$%
\end{tabular}
\right)
\]
and
\[
\psi(\rho,x)=\left(
\begin{tabular}
[c]{c}%
$\psi_{1}(\rho,x)$\\
$\psi_{2}(\rho,x)$%
\end{tabular}
\right)  ,
\]
which satisfy the asymptotic relations
\begin{equation}
\varphi(\rho,x)\sim\left(
\begin{tabular}
[c]{c}%
$1$\\
$0$%
\end{tabular}
\right)  e^{-i\rho x},\quad x\rightarrow-\infty\label{phi asympt}%
\end{equation}
and
\begin{equation}
\psi(\rho,x)\sim\left(
\begin{tabular}
[c]{c}%
$0$\\
$1$%
\end{tabular}
\right)  e^{i\rho x},\quad x\rightarrow\infty. \label{psi asympt}%
\end{equation}

Solutions $\varphi(\rho,x)$ and $\widetilde{\varphi}(\rho,x)$ as well as
$\psi(\rho,x)$ and $\widetilde{\psi}(\rho,x)$ are linearly independent, and
$W\left[  \widetilde{\varphi}(\rho,x),\varphi(\rho,x)\right]  =W\left[
\widetilde{\psi}(\rho,x),\psi(\rho,x)\right]  =1$. Thus, for any $\rho
\in\overline{\mathbb{C}^{+}}$, there exist the scalars $c_{ij}(\rho)$,
$i,j\in\left\{  0,1\right\}  $, such that
\[
\varphi(\rho,x)=c_{11}(\rho)\psi(\rho,x)+c_{12}(\rho)\widetilde{\psi}%
(\rho,x),
\]%
\[
\psi(\rho,x)=c_{21}(\rho)\widetilde{\varphi}(\rho,x)+c_{22}(\rho)\varphi
(\rho,x).
\]
Moreover, $c_{21}(\rho)=-c_{12}(\rho)$ and $c_{22}(\rho)=c_{11}(-\rho)$,
$\rho\in\mathbb{R}$ (see, e.g., \cite[p. 72]{Lamb}). The notations are
frequently used%
\[
\mathbf{a}(\rho)=c_{12}(\rho)\quad\text{and}\quad\mathbf{b}(\rho)=c_{11}%
(\rho),
\]
and $\mathbf{a}(\rho)$, $\mathbf{b}(\rho)$ are called the scattering
coefficients or the nonlinear Fourier coefficients (see, e.g.,
\cite{ChernyavskyFrumin2024JIIP}, \cite{Yousefi et al 2014 IEEE}). The
relation holds \cite[p. 73]{Lamb}%
\begin{equation}
\left\vert \mathbf{a}(\rho)\right\vert ^{2}+\left\vert \mathbf{b}%
(\rho)\right\vert ^{2}=1,\quad\rho\in\mathbb{R}.
\label{sum of modules squared =1}%
\end{equation}
The following equalities are valid
\[
\mathbf{b}(\rho)=W\left[  \widetilde{\psi}(\rho,x),\varphi(\rho,x)\right]  ,
\]%
\[
\mathbf{a}(\rho)=W\left[  \varphi(\rho,x),\psi(\rho,x)\right]  .
\]
Often the reflection and transmission coefficients are considered, which are
introduced as
\[
R(\rho):=\frac{\mathbf{b}(\rho)}{\mathbf{a}(\rho)}\quad\text{and}\quad
T(\rho):=\frac{1}{\mathbf{a}(\rho)}.
\]
Poles of $T(\rho)$ or the zeros of $\mathbf{a}(\rho)$ determine the
eigenvalues of (\ref{ZS1}), (\ref{ZS2}). For each eigenvalue $\rho_{m}$ there
is a norming constant $\mathbf{c}(\rho_{m})$, which links the Jost solutions
by
\begin{equation}
\varphi(\rho_{m},x)=\mathbf{c}(\rho_{m})\psi(\rho_{m},x). \label{phi = c psi}%
\end{equation}

The scattering coefficients, the eigenvalues and the norming constants
comprise the scattering data
\begin{equation}
SD:=\left\{  \mathbf{a}(\rho),\,\mathbf{b}(\rho),\,\rho\in\mathbb{R};\left\{
\rho_{m},\mathbf{c}(\rho_{m})\right\}  _{m=1}^{M}\right\}  . \label{SD}%
\end{equation}

Note that often, the single reflection coefficient $R(\rho)$ is icluded in the
set of the scattering data, instead of two coefficients $\mathbf{a}(\rho)$
and$\,\mathbf{b}(\rho),\,\rho\in\mathbb{R}$. However, it is mainly related
with the construction of the integral equations for solving the inverse
scattering problem. Application of the inverse scattering transform method
involving the Zakharov-Shabat system goes equally well with the scattering
data in the form (\ref{SD}), since the evolution laws are available for both
$\mathbf{a}(\rho)$ and$\,\mathbf{b}(\rho)$.

Given a real valued, piecewise continuously differentiable potential $q(x)$
decaying sufficiently fast when $\left\vert x\right\vert \rightarrow\infty$,
the direct scattering problem consists in finding the set of the scattering
data (\ref{SD}). The inverse scattering problem consists in recovering $q(x)$
from the scattering data.

\section{Spectral parameter power series representations for Jost solutions}

\subsection{Relation to a couple of Schr\"{o}dinger equations}

Let the pair $n_{1}(x)$ and $n_{2}(x)$ be a solution of (\ref{ZS1}),
(\ref{ZS2}) for a fixed $\rho$. Consider the functions%
\[
u=n_{1}+in_{2}\quad\text{and}\quad v=n_{1}-in_{2}.
\]
It is easy to derive from (\ref{ZS1}), (\ref{ZS2}) that $u$ and $v$ are
solutions of the second order equations%
\[
-\left(  \frac{d}{dx}-iq(x)\right)  \left(  \frac{d}{dx}+iq(x)\right)
u(x)=\rho^{2}u(x)
\]
and
\[
-\left(  \frac{d}{dx}+iq(x)\right)  \left(  \frac{d}{dx}-iq(x)\right)
v(x)=\rho^{2}v(x),
\]
respectively. The equations can be written in the form%
\begin{equation}
-u^{\prime\prime}(x)+q_{1}(x)u(x)=\rho^{2}u(x) \label{Schrod q1}%
\end{equation}
and
\begin{equation}
-v^{\prime\prime}(x)+q_{2}(x)v(x)=\rho^{2}v(x), \label{Schrod q2}%
\end{equation}
where
\begin{equation}
q_{1}(x):=-iq^{\prime}(x)-q^{2}(x)\quad\text{and}\quad q_{2}(x):=iq^{\prime
}(x)-q^{2}(x). \label{q1 and q2}%
\end{equation}

Let us use this relation between system (\ref{ZS1}), (\ref{ZS2}) and
Schr\"{o}dinger equations (\ref{Schrod q1}), (\ref{Schrod q2}) for
constructing series representations for the Jost solutions $\varphi(\rho,x)$
and $\psi(\rho,x)$. Consider
\begin{equation}
u_{L}(\rho,x):=\varphi_{1}(\rho,x)+i\varphi_{2}(\rho,x) \label{u_L}%
\end{equation}
and
\begin{equation}
v_{L}(\rho,x):=\varphi_{1}(\rho,x)-i\varphi_{2}(\rho,x), \label{v_L}%
\end{equation}
where $\varphi_{1}(\rho,x)$ and $\varphi_{2}(\rho,x)$ are the components of
the Jost solution $\varphi(\rho,x)$. We have that $u_{L}(\rho,x)$ and
$v_{L}(\rho,x)$ are solutions of (\ref{Schrod q1}) and (\ref{Schrod q2}),
respectively, and due to (\ref{phi asympt}) they satisfy the asymptotic
relations $u_{L}(\rho,x)\sim e^{-i\rho x}$, $v_{L}(\rho,x)\sim e^{-i\rho x}$,
when $x\rightarrow-\infty$. In other words $u_{L}(\rho,x)$ and $v_{L}(\rho,x)$
are the \textquotedblleft left\textquotedblright\ Jost solutions for
(\ref{Schrod q1}) and (\ref{Schrod q2}), respectively.

Analogously, the functions
\begin{equation}
u_{R}(\rho,x):=\psi_{2}(\rho,x)-i\psi_{1}(\rho,x)\sim e^{i\rho x},\quad
x\rightarrow\infty. \label{u_R}%
\end{equation}
and
\begin{equation}
v_{R}(\rho,x):=\psi_{2}(\rho,x)+i\psi_{1}(\rho,x)\sim e^{i\rho x},\quad
x\rightarrow\infty. \label{v_R}%
\end{equation}
are the \textquotedblleft right\textquotedblright\ Jost solutions for
(\ref{Schrod q1}) and (\ref{Schrod q2}), respectively.

Note that
\begin{equation}
\varphi_{1}(\rho,x)=\frac{1}{2}\left(  u_{L}(\rho,x)+v_{L}(\rho,x)\right)
,\quad\varphi_{2}(\rho,x)=-\frac{i}{2}\left(  u_{L}(\rho,x)-v_{L}%
(\rho,x)\right)  , \label{phi12 in terms of uL vL}%
\end{equation}%
\begin{equation}
\psi_{1}(\rho,x)=\frac{i}{2}\left(  u_{R}(\rho,x)-v_{R}(\rho,x)\right)
,\quad\psi_{2}(\rho,x)=\frac{1}{2}\left(  u_{R}(\rho,x)+v_{R}(\rho,x)\right)
. \label{psi12 in terms of uR vR}%
\end{equation}

\subsection{SPPS for Jost solutions of Schr\"{o}dinger
equation\label{SPPS for Jost Schr}}

Consider the Schr\"{o}dinger equation
\begin{equation}
-y^{\prime\prime}(x)+Q(x)y(x)=\rho^{2}y(x),\quad x\in(-\infty,\infty).
\label{Schr}%
\end{equation}
We suppose that $Q(x)$ is a complex valued function satisfying the condition
\[
\int_{-\infty}^{\infty}(1+\left\vert x\right\vert )\left\vert Q(x)\right\vert
dx<\infty.
\]
In this case $Q(x)$ is said to belong to the class $\mathcal{L}_{1,1}%
(-\infty,\infty)$. Then (\ref{Schr}) possesses the unique Jost solutions
$e(\rho,x)$ and $g(\rho,x)$ satisfying the asymptotic relations
\[
e(\rho,x)\sim e^{i\rho x},\quad x\rightarrow\infty\qquad\text{and}\qquad
g(\rho,x)\sim e^{-i\rho x},\quad x\rightarrow-\infty,
\]
respectively. They admit \cite[Chapter 1]{LevitanInverse} the integral
representations (sometimes called Levin's representations \cite[Ch. 5, Sect.
1]{Chadan})%
\begin{equation}
e(\rho,x)=e^{i\rho x}+\int_{x}^{\infty}A(x,t)e^{i\rho t}dt\qquad
\text{and}\qquad g(\rho,x)=e^{-i\rho x}+\int_{-\infty}^{x}B(x,t)e^{-i\rho t}dt
\label{Levin rep}%
\end{equation}
where $A(x,\cdot)\in L_{2}\left(  x,\infty\right)  $, $B(x,\cdot)\in
L_{2}\left(  -\infty,x\right)  $ and
\begin{equation}
A(x,x)=\frac{1}{2}\int_{x}^{\infty}Q(t)dt\qquad\text{and}\qquad B(x,x)=\frac
{1}{2}\int_{-\infty}^{x}Q(t)dt. \label{AB(x,x)}%
\end{equation}

\begin{theorem}
\cite{Kr2019MMAS InverseAxis}, \cite{KrBook2020} The functions $A$ and $B$
admit the following series representations%
\begin{equation}
A(x,t)=\sum_{n=0}^{\infty}a_{n}(x)L_{n}(t-x)e^{\frac{x-t}{2}} \label{A series}%
\end{equation}
and
\begin{equation}
B(x,t)=\sum_{n=0}^{\infty}b_{n}(x)L_{n}(x-t)e^{-\frac{x-t}{2}}
\label{B series}%
\end{equation}
where $L_{n}$ stands for the Laguerre polynomial of order $n$.

For any $x\in\mathbb{R}$ fixed, the series converge in the norm of
$L_{2}\left(  x,\infty\right)  $ and $L_{2}\left(  -\infty,x\right)  $, respectively.
\end{theorem}

The Fourier-Laguerre coefficients $a_{n}(x)$, $b_{n}(x)$, $n=0,1,\ldots$ can
be constructed by the following recurrent integration procedure \cite{DKK
Jost} (see also \cite{KrBook2020}). In the first step the Jost solutions
$e\left(  \frac{i}{2},x\right)  $ and $g\left(  \frac{i}{2},x\right)  $ of
(\ref{Schr}) for $\rho=\frac{i}{2}$ need to be computed. Then%
\begin{equation}
a_{0}(x)=e\left(  \frac{i}{2},x\right)  e^{\frac{x}{2}}-1 \label{a0}%
\end{equation}
and
\begin{equation}
b_{0}(x)=g\left(  \frac{i}{2},x\right)  e^{-\frac{x}{2}}-1. \label{b0}%
\end{equation}
Second, the solutions%

\begin{equation}
\eta(x):=e\left(  \frac{i}{2},x\right)  \int_{0}^{x}\frac{dt}{e^{2}(\frac
{i}{2},t)} \label{Abel formula}%
\end{equation}
and
\begin{equation}
\xi(x):=g\left(  \frac{i}{2},x\right)  \int_{x}^{0}\frac{dt}{g^{2}(\frac{i}%
{2},t)} \label{Abel formula 2}%
\end{equation}
are computed.

Finally, the subsequent coefficients $a_{n}(x)$, $b_{n}(x)$ are computed as
follows
\begin{align}
a_{n}(x)  &  =a_{0}(x)-2e^{\frac{x}{2}}\left(  \eta(x)J_{1,n}(x)-e\left(
\frac{i}{2},x\right)  J_{2,n}(x)\right)  ,\nonumber\\
b_{n}(x)  &  =b_{0}(x)+2e^{-\frac{x}{2}}\left(  \xi(x)I_{1,n}(x)-g\left(
\frac{i}{2},x\right)  I_{2,n}(x)\right)  ,\nonumber
\end{align}
where
\begin{equation}
J_{1,n}(x)=J_{1,n-1}(x)-e^{-\frac{x}{2}}e\left(  \frac{i}{2},x\right)
a_{n-1}(x)-\int_{x}^{\infty}\left(  e\left(  \frac{i}{2},t\right)
e^{-\frac{t}{2}}\right)  ^{\prime}a_{n-1}(t)dt, \label{J1 recurrent}%
\end{equation}%
\begin{equation}
J_{2,n}(x)=J_{2,n-1}(x)-e^{-\frac{x}{2}}\eta(x)a_{n-1}(x)-\int_{x}^{\infty
}\left(  \eta(t)e^{-\frac{t}{2}}\right)  ^{\prime}a_{n-1}(t)dt
\label{J2 recurrent}%
\end{equation}%
\begin{equation}
I_{1,n}(x)=I_{1,n-1}(x)+e^{\frac{x}{2}}g\left(  \frac{i}{2},x\right)
b_{n-1}(x)-\int_{-\infty}^{x}\left(  g\left(  \frac{i}{2},t\right)
e^{\frac{t}{2}}\right)  ^{\prime}b_{n-1}(t)dt, \label{I1}%
\end{equation}
and
\begin{equation}
I_{2,n}(x)=I_{2,n-1}(x)+e^{\frac{x}{2}}\xi(x)b_{n-1}(x)-\int_{-\infty}%
^{x}\left(  \xi(t)e^{\frac{t}{2}}\right)  ^{\prime}b_{n-1}(t)dt, \label{I2}%
\end{equation}
assuming $J_{1,0}(x)=J_{2,0}(x)=I_{1,0}(x)=I_{2,0}(x)\equiv0$.

Thus, equation (\ref{Schr}) needs to be solved for a single value of the
parameter: $\rho=\frac{i}{2}$. For this, any numerical method can be used,
including the SPPS method (see e.g., \cite{KrBook2020}), as it is done in the
present work. This recurrent integration procedure allows one to compute
dozens of the Fourier-Laguerre coefficients $a_{n}(x)$ and $b_{n}(x)$.

Note that due to the identity $L_{n}(0)=1$, for all $n=0,1,\ldots$, from
(\ref{A series}) and (\ref{B series}) two useful relations follow%
\begin{equation}
\sum_{n=0}^{\infty}a_{n}(x)=A(x,x)=\frac{1}{2}\int_{x}^{\infty}q(t)dt,\quad
\sum_{n=0}^{\infty}b_{n}(x)=B(x,x)=\frac{1}{2}\int_{-\infty}^{x}q(t)dt.
\label{sumab}%
\end{equation}

Denote
\begin{equation}
z=z\left(  \rho\right)  :=\frac{\frac{1}{2}+i\rho}{\frac{1}{2}-i\rho}.
\label{z}%
\end{equation}
Notice that this is a M\"{o}bius transformation of the upper halfplane of the
complex variable $\rho$ onto the unit disc $D=\left\{  z\in\mathbb{C}%
:\,\left\vert z\right\vert \leq1\right\}  $.

Substitution of (\ref{A series}) and (\ref{B series}) into (\ref{Levin rep})
leads to the following series representations for the Jost solutions in terms
of the parameter $z$ \cite{Kr2019MMAS InverseAxis}, \cite{KrBook2020}%
\begin{equation}
e(\rho,x)=e^{i\rho x}\left(  1+\left(  z+1\right)  \sum_{n=0}^{\infty}\left(
-1\right)  ^{n}z^{n}a_{n}(x)\right)  \label{e via z}%
\end{equation}
and
\begin{equation}
g(\rho,x)=e^{-i\rho x}\left(  1+\left(  z+1\right)  \sum_{n=0}^{\infty}\left(
-1\right)  ^{n}z^{n}b_{n}(x)\right)  . \label{g via z}%
\end{equation}
Note that the functions $e(\rho,x)e^{-i\rho x}$ and $g(\rho,x)e^{i\rho x}$ are
represented as power series with respect to the parameter $z$. For any
$x\in\mathbb{R}$ the series $\sum_{n=0}^{\infty}\left\vert a_{n}(x)\right\vert
^{2}$ and $\sum_{n=0}^{\infty}\left\vert b_{n}(x)\right\vert ^{2}$ converge,
which is a consequence of the fact that they are Fourier coefficients with
respect to the system of Laguerre polynomials of corresponding functions from
$L_{2}\left(  0,\infty;e^{-t}\right)  $. Hence for any $x\in\mathbb{R}$ the
functions $e(\rho,x)e^{-i\rho x}$ and $g(\rho,x)e^{i\rho x}$ belong to the
Hardy space $H^{2}(D)$ as functions of $z$ (this is due to the well known
result from complex analysis, see, e.g., \cite[Theorem 17.12]{Rudin}).

We call (\ref{e via z}) and (\ref{g via z}) spectral parameter power series
(SPPS) representations for the Jost solutions.

\subsection{SPPS for Jost solutions of Zakharov-Shabat
system\label{subsect SPPS for ZS}}

The "left" and "right" Jost solutions $u_{L}(\rho,x)$, $u_{R}(\rho,x)$ of
equation (\ref{Schrod q1}) and $v_{L}(\rho,x)$, $v_{R}(\rho,x)$ of equation
(\ref{Schrod q2}) admit their respective SPPS representations of the type
(\ref{e via z}) and (\ref{g via z}). We have
\[
u_{L}(\rho,x)=e^{-i\rho x}(1+\left(  z+1\right)  \sum_{n=0}^{\infty}\left(
-1\right)  ^{n}z^{n}b_{1,n}(x)),\quad u_{R}(\rho,x)=e^{i\rho x}(1+\left(
z+1\right)  \sum_{n=0}^{\infty}\left(  -1\right)  ^{n}z^{n}a_{1,n}(x)),
\]%
\[
v_{L}(\rho,x)=e^{-i\rho x}(1+\left(  z+1\right)  \sum_{n=0}^{\infty}\left(
-1\right)  ^{n}z^{n}b_{2,n}(x)),\quad v_{R}(\rho,x)=e^{i\rho x}(1+\left(
z+1\right)  \sum_{n=0}^{\infty}\left(  -1\right)  ^{n}z^{n}a_{2,n}(x)),
\]
where the first subindex in the coefficients refers to the potential
$q_{1}(x)$ or $q_{2}(x)$, respectively. Let us show, however, that
\begin{equation}
b_{2,n}(x)=\overline{b}_{1,n}(x)\quad\text{and}\quad a_{2,n}(x)=\overline
{a}_{1,n}(x).\label{equality of coefs}%
\end{equation}
This is a consequence of the fact that $q_{2}(x)=\overline{q}_{1}(x)$. Indeed,
a function $y(x)$ is a solution of (\ref{Schrod q2}) with $\rho\in\mathbb{R}$
iff $\overline{y}(x)$ is a solution of equation (\ref{Schrod q1}) with the
same $\rho$. So, if $y(x)\sim e^{i\rho x}$, when $x\rightarrow\infty$, then
$\overline{y}(x)\sim e^{-i\rho x}$. This leads to the conclusion that
\[
u_{L}(\rho,x)=\overline{v}_{L}(-\rho,x)\quad\text{and}\quad u_{R}%
(\rho,x)=\overline{v}_{R}(-\rho,x).
\]
Note that $\overline{z}(-\rho)=z(\rho)$, $\rho\in\mathbb{R}$. Hence $\left(
z+1\right)  \sum_{n=0}^{\infty}\left(  -1\right)  ^{n}z^{n}b_{1,n}(x)=\left(
z+1\right)  \sum_{n=0}^{\infty}\left(  -1\right)  ^{n}z^{n}\overline{b}%
_{2,n}(x)$ for all $z\in\partial D$. From this equality of the convergent
power series we obtain the pairwise equality of their coefficients,
$b_{2,n}(x)=\overline{b}_{1,n}(x)$. Analogously, the second equality in
(\ref{equality of coefs}) is proved.

Thus, it is convenient to simplify the notation: $a_{n}(x)=a_{1,n}(x)$ and
$b_{n}(x)=b_{1,n}(x)$.

\begin{theorem}
\label{Th SPPS for ZS}Let the potential $q(x)$ in (\ref{ZS1}), (\ref{ZS2}) be
real valued, and such that $q^{2}(x)$, $q^{\prime}(x)\in\mathcal{L}%
_{1,1}(-\infty,\infty)$. Then the components of the Jost solutions
$\varphi(\rho,x)$ and $\psi(\rho,x)$ admit the following series
representations in terms of the spectral parameter $z=\frac{\frac{1}{2}+i\rho
}{\frac{1}{2}-i\rho}$:%
\begin{equation}
\varphi_{1}(\rho,x)=e^{-i\rho x}\left(  1+\left(  z+1\right)  \sum
_{n=0}^{\infty}\left(  -1\right)  ^{n}z^{n}\operatorname{Re}b_{n}(x)\right)
\label{phi1}%
\end{equation}
and
\begin{equation}
\varphi_{2}(\rho,x)=e^{-i\rho x}\left(  z+1\right)  \sum_{n=0}^{\infty}\left(
-1\right)  ^{n}z^{n}\operatorname{Im}b_{n}(x). \label{phi2}%
\end{equation}
Analogously,
\begin{equation}
\psi_{1}(\rho,x)=-e^{i\rho x}\left(  z+1\right)  \sum_{n=0}^{\infty}\left(
-1\right)  ^{n}z^{n}\operatorname{Im}a_{n}(x) \label{psi1}%
\end{equation}
and
\begin{equation}
\psi_{2}(\rho,x)=e^{i\rho x}\left(  1+\left(  z+1\right)  \sum_{n=0}^{\infty
}\left(  -1\right)  ^{n}z^{n}\operatorname{Re}a_{n}(x)\right)  , \label{psi2}%
\end{equation}
where $\left\{  a_{n}(x)\right\}  _{n=0}^{\infty}$ and $\left\{
b_{n}(x)\right\}  _{n=0}^{\infty}$ are the coefficients from (\ref{e via z})
and (\ref{g via z}), respectively, and for all $x\in(-\infty,\infty)$ the
power series with respect to $z$ converge in the unit disk $\left\vert
z\right\vert <1$. Moreover, since $\sum_{n=0}^{\infty}\left\vert
\operatorname{Re}a_{n}(x)\right\vert ^{2}$, $\sum_{n=0}^{\infty}\left\vert
\operatorname{Im}a_{n}(x)\right\vert ^{2}$, $\sum_{n=0}^{\infty}\left\vert
\operatorname{Re}b_{n}(x)\right\vert ^{2}$ and $\sum_{n=0}^{\infty}\left\vert
\operatorname{Im}b_{n}(x)\right\vert ^{2}$ converge, the functions
$\varphi_{1,2}(\rho,x)e^{i\rho x}$ and $\psi_{1,2}(\rho,x)e^{-i\rho x}$ belong
to the Hardy space $H^{2}(D)$ as functions of $z$.
\end{theorem}

\textbf{Proof. }From (\ref{phi12 in terms of uL vL}) and
(\ref{psi12 in terms of uR vR}) and taking into account
(\ref{equality of coefs}) we obtain the series representations (\ref{phi1}%
)-(\ref{psi2}). The results on the convergence of the series follow directly
from the corresponding results from subsection \ref{SPPS for Jost Schr}.

\begin{theorem}
\label{Th estimates}Let the potential $q(x)$ satisfy the conditions of Theorem
\ref{Th SPPS for ZS}. Then the remainders of the partial sums of the series
(\ref{phi1})-(\ref{psi2}) admit the following estimates.

(1) If $\operatorname{Im}\rho>0$, then
\begin{equation}
\left\vert \varphi_{1}(\rho,x)-\varphi_{1,N}(\rho,x)\right\vert \leq
\varepsilon_{N}(x)\frac{e^{-\operatorname{Im}\rho\,x}}{\sqrt
{2\operatorname{Im}\rho}} \label{estimate rho complex}%
\end{equation}
where
\begin{equation}
\varepsilon_{N}(x):=\left(  \sum_{N+1}^{\infty}\left\vert b_{n}(x)\right\vert
^{2}\right)  ^{1/2}. \label{from Parseval}%
\end{equation}

(2) For $\rho\in\mathbb{R}$,
\begin{equation}
\left\Vert \varphi_{1}(\cdot,x)-\varphi_{1,N}(\cdot,x)\right\Vert
_{L_{2}(-\infty,\infty)}\leq\sqrt{2\pi}\varepsilon_{N}(x).
\label{estimate rho real}%
\end{equation}

Analogous inequalities are valid for the partial sums of the series
(\ref{phi2})-(\ref{psi2}).
\end{theorem}

\textbf{Proof. }The proof follows directly from the analogous result for the
series representations of the Jost solutions of the Schr\"{o}dinger equation,
see \cite[Theorem 10.2]{KrBook2020}. For example, to obtain
(\ref{estimate rho complex}) we observe that
\begin{align*}
\left\vert \varphi_{1}(\rho,x)-\varphi_{1,N}(\rho,x)\right\vert  &  \leq
\frac{1}{2}\left(  \left\vert u_{L}(\rho,x)-u_{L,N}(\rho,x)\right\vert
+\left\vert v_{L}(\rho,x)-v_{L,N}(\rho,x)\right\vert \right)  \\
&  \leq\varepsilon_{N}(x)\frac{e^{-\operatorname{Im}\rho\,x}}{\sqrt
{2\operatorname{Im}\rho}},
\end{align*}
where we made use of the estimates for $u_{L}(\rho,x)$ and $v_{L}(\rho,x)$, as
well as of the fact that the SPPS coefficients of $u_{L}(\rho,x)$ and
$v_{L}(\rho,x)$ are mutually conjugate, and hence $\varepsilon_{N}(x)$ is the
same for both functions. The case of real $\rho$ is considered similarly.

From (\ref{estimate rho complex}) we notice that for larger $\operatorname{Im}%
\rho$ the convergence is faster.

\begin{theorem}
\label{Th q from b0 and a0}The following equalities hold
\begin{align}
q(x)  &  =\frac{(e^{\frac{x}{2}}\left(  1+\operatorname{Re}b_{0}%
(x)+\operatorname{Im}b_{0}(x)\right)  )^{\prime}}{e^{\frac{x}{2}}\left(
\operatorname{Im}b_{0}(x)-\operatorname{Re}b_{0}(x)-1\right)  }+\frac{1}%
{2}\nonumber\\
&  =\frac{\frac{1}{2}(1+\operatorname{Re}b_{0}(x)+\operatorname{Im}%
b_{0}(x))+\operatorname{Re}b_{0}^{\prime}(x)+\operatorname{Im}b_{0}^{\prime
}(x)}{\operatorname{Im}b_{0}(x)-\operatorname{Re}b_{0}(x)-1}+\frac{1}{2}
\label{q from b0}%
\end{align}
and
\begin{align}
q(x)  &  =\frac{(e^{-\frac{x}{2}}\left(  1+\operatorname{Re}a_{0}%
(x)-\operatorname{Im}a_{0}(x)\right)  )^{\prime}}{e^{-\frac{x}{2}}\left(
1+\operatorname{Re}a_{0}(x)+\operatorname{Im}a_{0}(x)\right)  }+\frac{1}%
{2}\nonumber\\
&  =\frac{-\frac{1}{2}(1+\operatorname{Re}a_{0}(x)-\operatorname{Im}%
a_{0}(x))+\operatorname{Re}a_{0}^{\prime}(x)-\operatorname{Im}a_{0}^{\prime
}(x)}{1+\operatorname{Re}a_{0}(x)+\operatorname{Im}a_{0}(x)}+\frac{1}{2}.
\label{q from a0}%
\end{align}

\end{theorem}

\textbf{Proof. }Indeed, consider Zakharov-Shabat system (\ref{ZS1}),
(\ref{ZS2}) for $\rho=\frac{i}{2}$. We have
\begin{equation}
\frac{dn_{1}(x)}{dx}-\frac{n_{1}(x)}{2}=q(x)n_{2}(x), \label{ZS1i/2}%
\end{equation}%
\begin{equation}
\frac{dn_{2}(x)}{dx}+\frac{n_{2}(x)}{2}=-q(x)n_{1}(x). \label{ZS2i/2}%
\end{equation}
It is easy to see that if a nontrivial solution $n_{1}(x)$, $n_{2}(x)$ is
known, $q(x)$ can be recovered, e.g., from the equality%
\begin{equation}
q(x)=\frac{\left(  n_{1}(x)+n_{2}(x)\right)  ^{\prime}}{n_{2}(x)-n_{1}%
(x)}+\frac{1}{2}. \label{q in terms of n1 n2}%
\end{equation}
Note that $z\left(  \frac{i}{2}\right)  =0$, and hence the components of the
Jost solutions for (\ref{ZS1i/2}), (\ref{ZS2i/2}) have the form
\[
\varphi_{1}(\frac{i}{2},x)=e^{\frac{x}{2}}\left(  1+\operatorname{Re}%
b_{0}(x)\right)  ,\quad\varphi_{2}(\frac{i}{2},x)=e^{\frac{x}{2}%
}\operatorname{Im}b_{0}(x),
\]%
\[
\psi_{1}(\frac{i}{2},x)=-e^{-\frac{x}{2}}\operatorname{Im}a_{0}(x),\quad
\psi_{2}(\frac{i}{2},x)=e^{-\frac{x}{2}}\left(  1+\operatorname{Re}%
a_{0}(x)\right)  .
\]
Thus, substituting $\varphi_{1}(\frac{i}{2},x)$ and $\varphi_{2}(\frac{i}%
{2},x)$ into (\ref{q in terms of n1 n2}) we obtain (\ref{q from b0}), and
analogously, substituting $\psi_{1}(\frac{i}{2},x)$ and $\psi_{2}(\frac{i}%
{2},x)$ into (\ref{q in terms of n1 n2}) we obtain (\ref{q from a0}).

This result shows that the knowledge of $a_{0}(x)$ or $b_{0}(x)$ is sufficient
for recovering $q(x)$.

\section{Solving the direct scattering problem\label{SectDirect}}

Let us obtain series expansions for the scattering coefficients $\mathbf{a}%
(\rho)$ and$\,\mathbf{b}(\rho)$. We have
\begin{align}
\mathbf{a}(\rho)  &  =W\left[  \varphi(\rho,x),\psi(\rho,x)\right]
=\varphi_{1}(\rho,0)\psi_{2}(\rho,0)-\varphi_{2}(\rho,0)\psi_{1}%
(\rho,0)\nonumber\\
&  =\left(  1+\left(  z+1\right)  \sum_{n=0}^{\infty}\left(  -1\right)
^{n}z^{n}\operatorname{Re}b_{n}(0)\right)  \left(  1+\left(  z+1\right)
\sum_{n=0}^{\infty}\left(  -1\right)  ^{n}z^{n}\operatorname{Re}%
a_{n}(0)\right) \nonumber\\
&  -\left(  z+1\right)  ^{2}\left(  \sum_{n=0}^{\infty}\left(  -1\right)
^{n}z^{n}\operatorname{Im}b_{n}(0)\right)  \left(  \sum_{n=0}^{\infty}\left(
-1\right)  ^{n}z^{n}\operatorname{Im}a_{n}(0)\right)  \label{a(rho)}%
\end{align}
for all $\rho\in\overline{\mathbb{C}^{+}}$ and
\begin{align}
\mathbf{b}(\rho)  &  =W\left[  \widetilde{\psi}(\rho,x),\varphi(\rho
,x)\right]  =\psi_{2}(-\rho,0)\varphi_{2}(\rho,0)+\psi_{1}(-\rho,0)\varphi
_{1}(\rho,0)\nonumber\\
&  =\left(  1+\left(  \overline{z}+1\right)  \sum_{n=0}^{\infty}\left(
-1\right)  ^{n}\overline{z}^{n}\operatorname{Re}a_{n}(0)\right)  \left(
\left(  z+1\right)  \sum_{n=0}^{\infty}\left(  -1\right)  ^{n}z^{n}%
\operatorname{Im}b_{n}(0)\right) \nonumber\\
&  -\left(  \overline{z}+1\right)  \left(  \sum_{n=0}^{\infty}\left(
-1\right)  ^{n}\overline{z}^{n}\operatorname{Im}a_{n}(0)\right)  \left(
1+\left(  z+1\right)  \sum_{n=0}^{\infty}\left(  -1\right)  ^{n}%
z^{n}\operatorname{Re}b_{n}(0)\right)  \label{b(rho)}%
\end{align}
for all $\rho\in\mathbb{R}$. These formulas allow us to compute the scattering
data (\ref{SD}).

Note that from (\ref{a(rho)}) the well known parity of eigenvalues can be
derived: $\rho_{m}$ and $-\overline{\rho_{m}}$ can be zeros of $\mathbf{a}%
(\rho)$ only simultaneously. This is because, obviously, $\mathbf{a}(\rho)=0$
iff $\overline{\mathbf{a}(\rho)}=0$, that is the expression (\ref{a(rho)})
with $z\ $replaced by $\overline{z}$ equals zero, and at the same time:
$\overline{z}(\rho)=z(-\overline{\rho})$. More generally, $\overline
{\mathbf{a}(\rho)}=\mathbf{a}(-\overline{\rho})$ for all $\rho\in
\overline{\mathbb{C}^{+}\text{,}}$ and $\overline{\mathbf{b}}(\rho
)=\mathbf{b}(-\rho)$ for all $\rho\in\mathbb{R}$.

\begin{remark}
\label{Rem choice of N}For computations, series in (\ref{a(rho)}) and
(\ref{b(rho)}) need to be truncated up to certain $N\in\mathbb{N}$. The choice
of its convenient value can be made by using equalities (\ref{sumab}). Indeed,
considering
\[
\varepsilon_{L}(N):=\left\vert \sum_{n=0}^{N}b_{n}(0)-\frac{1}{2}\int
_{-\infty}^{0}q_{1}(t)dt\right\vert
\]
and
\[
\varepsilon_{R}(N):=\left\vert \sum_{n=0}^{N}a_{n}(0)-\frac{1}{2}\int
_{0}^{\infty}q_{1}(t)dt\right\vert
\]
one can look for the values $N_{L}$ and $N_{R}$ delivering $\min
_{N}\varepsilon_{L}$ and $\min_{N}\varepsilon_{R}$. In our numerical examples,
for simplicity, we chose the same value for left and right, taken as the
biggest of $N_{L}$ and $N_{R}$.
\end{remark}

\begin{remark}
\label{Rem eigenvalues}The computation of the eigenvalues reduces to the
location of zeros of the expression (\ref{a(rho)}) inside of the unit disk
$D$. For numerical implementation one considers truncated sums in
(\ref{a(rho)}), and hence the computation of the eigenvalues reduces to the
location of roots of a polynomial of the variable $z$ in the interior of the
unit disk $D$.

The norming constants are computed from (\ref{phi = c psi}) as follows%
\begin{equation}
\mathbf{c}(\rho_{m})=\frac{\varphi_{1}(\rho_{m},0)}{\psi_{1}(\rho_{m}%
,0)}=-\frac{\left(  1+\left(  z_{m}+1\right)  \sum_{n=0}^{\infty}\left(
-1\right)  ^{n}z_{m}^{n}\operatorname{Re}b_{n}(0)\right)  }{\left(
z_{m}+1\right)  \sum_{n=0}^{\infty}\left(  -1\right)  ^{n}z_{m}^{n}%
\operatorname{Im}a_{n}(0)}, \label{c(rho)}%
\end{equation}
where $z_{m}=z(\rho_{m})$. Note that $\mathbf{c}(-\overline{\rho_{m}%
})=\overline{\mathbf{c}(\rho_{m})}$. For numerical implementation, again, one
considers truncated sums in (\ref{c(rho)}).
\end{remark}

Summarizing, solution of the direct scattering problem goes as follows. Given
the potential $q(x)$, compute a sufficiently large set of the coefficients
$\left\{  a_{n}(0),b_{n}(0)\right\}  _{n=0}^{N}$, following the recurrent
integration procedure above. Next, $\mathbf{a}(\rho)$ and$\,\mathbf{b}(\rho)$
for $\rho\in\mathbb{R}$ are computed from (\ref{a(rho)}) and (\ref{b(rho)}),
where the truncated sums up to $N$ are considered, and $z(\rho)\in\partial D$.
Finally, the eigenvalues and the norming constants are computed according to
Remark \ref{Rem eigenvalues}.

\section{Solving the inverse scattering problem}

Given the scattering data (\ref{SD}), we consider the equality%
\begin{equation}
\varphi(\rho,x)=\mathbf{b}(\rho)\psi(\rho,x)+\mathbf{a}(\rho)\widetilde{\psi
}(\rho,x),\quad\rho\in\mathbb{R} \label{phi = b psi +}%
\end{equation}
and, additionally, if the discrete spectrum is not empty, the equalities
(\ref{phi = c psi}) for all $\rho_{m}$, $m=1,\ldots,M$. This allows us to
construct a system of linear algebraic equations for the coefficients
$\left\{  a_{n}(x),b_{n}(x)\right\}  $ as follows. First of all, for every
$\rho\in\mathbb{R}$, equality (\ref{phi = b psi +}) written componentwise
gives us two equations%
\[
\varphi_{1}(\rho,x)=\mathbf{b}(\rho)\psi_{1}(\rho,x)+\mathbf{a}(\rho)\psi
_{2}(-\rho,x),
\]%
\[
\varphi_{2}(\rho,x)=\mathbf{b}(\rho)\psi_{2}(\rho,x)-\mathbf{a}(\rho)\psi
_{1}(-\rho,x).
\]
Substitution of the series from (\ref{phi1})-(\ref{psi2}) gives%
\begin{gather}
e^{-i\rho x}\left(  z+1\right)  \sum_{n=0}^{\infty}\left(  -1\right)
^{n}z^{n}\operatorname{Re}b_{n}(x)-\mathbf{a}(\rho)e^{-i\rho x}\left(
\overline{z}+1\right)  \sum_{n=0}^{\infty}\left(  -1\right)  ^{n}\overline
{z}^{n}\operatorname{Im}b_{n}(x)\nonumber\\
+\mathbf{b}(\rho)e^{i\rho x}\left(  z+1\right)  \sum_{n=0}^{\infty}\left(
-1\right)  ^{n}z^{n}\operatorname{Im}a_{n}(x)=\left(  \mathbf{a}%
(\rho)-1\right)  e^{-i\rho x}, \label{s1}%
\end{gather}%
\begin{gather}
e^{-i\rho x}\left(  z+1\right)  \sum_{n=0}^{\infty}\left(  -1\right)
^{n}z^{n}\operatorname{Im}b_{n}(x)-\mathbf{b}(\rho)e^{i\rho x}\left(
z+1\right)  \sum_{n=0}^{\infty}\left(  -1\right)  ^{n}z^{n}\operatorname{Re}%
a_{n}(x)\nonumber\\
-\mathbf{a}(\rho)e^{-i\rho x}\left(  \overline{z}+1\right)  \sum_{n=0}%
^{\infty}\left(  -1\right)  ^{n}\overline{z}^{n}\operatorname{Im}%
a_{n}(x)=\mathbf{b}(\rho)e^{i\rho x}. \label{s2}%
\end{gather}
Additionally, for every eigenvalue $\rho_{m}$ we have an equality
(\ref{phi = c psi}), which can be written in the form of the following two
scalar equations%

\begin{equation}
e^{-i\rho_{m}x}\left(  z_{m}+1\right)  \sum_{n=0}^{\infty}\left(  -1\right)
^{n}z_{m}^{n}\operatorname{Re}b_{n}(x)+\mathbf{c}(\rho_{m})e^{i\rho_{m}%
x}\left(  z_{m}+1\right)  \sum_{n=0}^{\infty}\left(  -1\right)  ^{n}z_{m}%
^{n}\operatorname{Im}a_{n}(x)=-e^{-i\rho_{m}x} \label{s3}%
\end{equation}
and%
\begin{equation}
e^{-i\rho_{m}x}\left(  z_{m}+1\right)  \sum_{n=0}^{\infty}\left(  -1\right)
^{n}z_{m}^{n}\operatorname{Im}b_{n}(x)-\mathbf{c}(\rho_{m})e^{i\rho_{m}%
x}\left(  z_{m}+1\right)  \sum_{n=0}^{\infty}\left(  -1\right)  ^{n}z_{m}%
^{n}\operatorname{Re}a_{n}(x)=\mathbf{c}(\rho_{m})e^{i\rho_{m}x}. \label{s4}%
\end{equation}

Equations (\ref{s1})-(\ref{s4}) lead to a system of linear algebraic equations
for the four sets of the coefficients $\left\{  \operatorname{Re}%
a_{n}(x),\,\operatorname{Im}a_{n}(x),\,\operatorname{Re}b_{n}%
(x),\,\operatorname{Im}b_{n}(x)\right\}  $ for every $x$. First of all we
notice that since the coefficients are real, each of the equations in fact
gives us two equations, by considering the real and imaginary parts of it
separately. Second, we choose a sufficiently large number $K$ of points
$\rho_{k}\in\mathbb{R}$\ (which are substituted into (\ref{s1}) and
(\ref{s2})), and finally we choose $N\in\mathbb{N}$ and consider the truncated
series in (\ref{s1})-(\ref{s4}) up to $N$. Thus, for every $x$ from an
interval of interest $x\in\left[  -a,a\right]  $, we can write an
overdetermined system of the form
\begin{equation}
AX=B, \label{main sys}%
\end{equation}
where
\begin{equation}
X=(\left\{  \operatorname{Re}b_{n}(x)\right\}  _{n=0}^{N},\left\{
\operatorname{Im}b_{n}(x)\right\}  _{n=0}^{N},\left\{  \operatorname{Re}%
a_{n}(x)\right\}  _{n=0}^{N},\left\{  \operatorname{Im}a_{n}(x)\right\}
_{n=0}^{N})^{T}, \label{X}%
\end{equation}
the matrix $A$ has the dimension $4\left(  K+M\right)  \times4(N+1)$; $4K$
equations are obtained from (\ref{s1}) and (\ref{s2}), each $\rho_{k}$,
$k=1,\ldots,K$ gives four equations (real and imaginary parts of (\ref{s1})
and (\ref{s2})), and $4M$ equations obtained from (\ref{s3}) and (\ref{s4}),
each eigenvalue $\rho_{m}$, $m=1,\ldots,M$ also gives four equations (real and
imaginary parts of (\ref{s3}) and (\ref{s4})). The vector of the right-hand
side $B$ is obtained from the right-hand side of (\ref{s1})-(\ref{s4}), correspondingly.

When solving the resulting system of equations, we are mostly interested in
the pair of the very first coefficients $\operatorname{Re}b_{0}(x)$,
$\operatorname{Im}b_{0}(x)$, or $\operatorname{Re}a_{0}(x)$,
$\operatorname{Im}a_{0}(x)$. According to Theorem \ref{Th q from b0 and a0},
their knowledge leads to the recovery of the potential $q(x)$.

Summarizing, the algorithm for solving the inverse scattering problem consists
of two steps. First, the system of linear algebraic equations (\ref{main sys})
is solved at a sufficiently dense set of points $x_{j}\in\left[  -a,a\right]
$ for some $a>0$. Second, $q(x)$ is recovered from (\ref{q from b0}) or
(\ref{q from a0}).

The choice of the parameter $N$ in (\ref{X}) can be performed by using, for
example, the fact that
\[
W\left[  \varphi(\frac{i}{2},x),\psi(\frac{i}{2},x)\right]  =\left(
1+\operatorname{Re}b_{0}(x)\right)  \left(  1+\operatorname{Re}a_{0}%
(x)\right)  +\operatorname{Im}b_{0}(x)\operatorname{Im}a_{0}(x)=\text{Const.}%
\]
This constant equals of course $\mathbf{a}(\frac{i}{2})$, which is unknown.
However, considering the expression
\[
\varepsilon(N)=\max_{x}\left\vert \frac{d}{dx}\left(  \left(
1+\operatorname{Re}b_{0}(x)\right)  \left(  1+\operatorname{Re}a_{0}%
(x)\right)  +\operatorname{Im}b_{0}(x)\operatorname{Im}a_{0}(x)\right)
\right\vert
\]
as a function of $N$, which is chosen in (\ref{X}) for computing $a_{0}(x)$
and $b_{0}(x)$ involved, one can look for the value of $N$ delivering the
minimum of $\varepsilon(N)$. This value is used then when formulating
(\ref{main sys}).

\section{Numerical examples}

The proposed approach can be implemented directly using an available numerical
computing environment. All the reported computations were performed in Matlab
R2024a on an Intel i7-1360P equipped laptop computer and took no more than
several seconds.

\textbf{Example 1. }Consider the potential \cite[pp. 75-77]{Lamb}
\[
q(x)=\mu\operatorname{sech}\left(  \mu x\right)  .
\]
This is a reflectionless potential, $\mathbf{b}(\rho)\equiv0$, $\rho\in\left(
-\infty,\infty\right)  $. It possesses one eigenvalue $\rho_{1}=\frac{i\mu}%
{2}$. 

For the numerical computation we chose $\mu=\pi$. The computation of the
truncation parameter $N$, performed following Remark \ref{Rem choice of N},
delivered $N=95$. With this choice of $N$, the eigenvalue was computed with
the absolute error $2.8\cdot10^{-15}$. We mention that for this, the location
of the roots of the polynomial of the variable $z$ (see Remark
\ref{Rem eigenvalues}) was performed with the aid of the Matlab routine
`roots'. The scattering coefficients $\mathbf{a}(\rho)$ and$\,\mathbf{b}%
(\rho)$ were computed following Section \ref{SectDirect} at $4000$ points
$\rho\in\left[  -30,30\right]  $. The computed scattering data were used as
the input data for the inverse scattering problem. 

Fig.1 presents the result of the recovery of the potential (\ref{q Ex2sech}).
\begin{figure}
[ptb]
\begin{center}
\includegraphics[
height=3.4506in,
width=4.593in
]%
{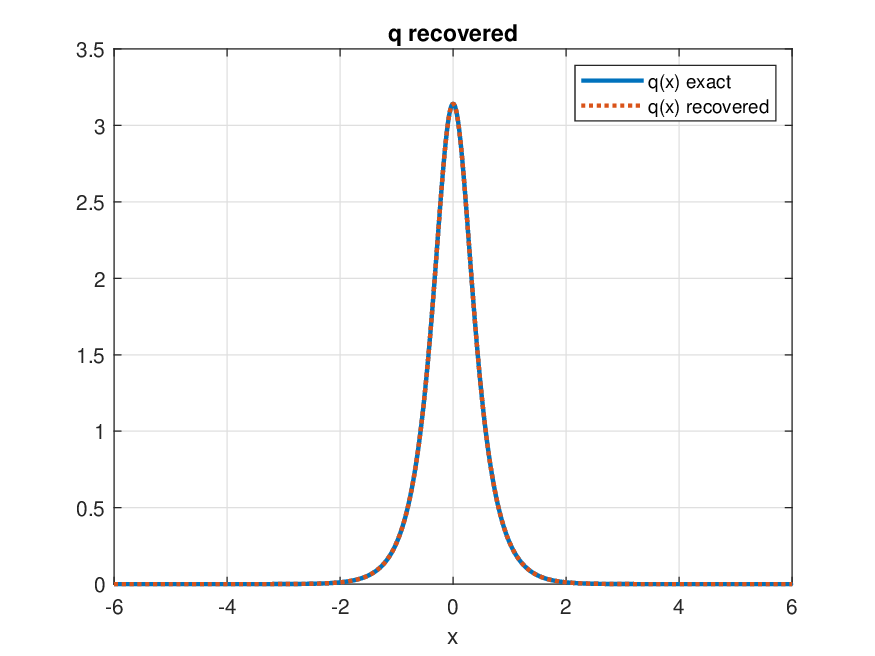}%
\caption{Potential from Example 1, recovered with the maximum absolute error
$1.3\cdot10^{-7}$.}%
\label{Fig1}%
\end{center}
\end{figure}
$N$ in (\ref{X}), which defines the number of the unknown coefficients in each
SPPS representation, resulted in $N=25$. One eigenvalue added $4$ equations,
obtained from (\ref{s3}) and (\ref{s4}). Thus, matrix $A$ in (\ref{main sys})
resulted of the dimension $16004\times104$. The maximum absolute error of the
recovered potential resulted in $1.3\cdot10^{-7}$. 

\bigskip

\textbf{Example 2. }Consider the potential \cite{Satsuma et al}
\begin{equation}
q(x)=\mu\operatorname{sech}x\label{q Ex2sech}%
\end{equation}
with the amplitude $\mu$ being positive. In this case the eigenvalues are
determined by the formula (see, e.g., \cite{Mullyadzhanov et al 2021})
\[
\rho_{m}=i\left(  \mu-m+\frac{1}{2}\right)  ,\quad m=1,\ldots,\left\lfloor
\mu+\frac{1}{2}\right\rfloor ,
\]
where $\left\lfloor \cdot\right\rfloor $ stands for the integer part, and the
closed form expressions for the scattering coefficients are available. In
particular, $\mathbf{b}(\rho)=-\frac{\sin(\pi\mu)}{\cosh(\pi\rho)}$.

For the numerical computation we chose $\mu=5+\frac{\pi}{7}$. Thus, $q(x)$
possesses five eigenvalues. The computation of the truncation parameter $N$,
performed following Remark \ref{Rem choice of N}, delivered $N=197$. With this
choice of $N$, the eigenvalues were computed with the maximum absolute error
(attained for that with the biggest absolute value) $4.5\cdot10^{-12}$. As in
other numerical examples, the location of the roots of the polynomial of the
variable $z$ (see Remark \ref{Rem eigenvalues}) was performed with the aid of
the Matlab routine `roots'. The scattering coefficients $\mathbf{a}(\rho)$
and$\,\mathbf{b}(\rho)$ were computed following Section \ref{SectDirect} at
$4000$ points $\rho\in\left[  -30,30\right]  $, and the maximum absolute error
of the computed $\mathbf{b}(\rho)$ resulted in $6.6\cdot10^{-13}$. The
computed scattering data were used as the input data for the inverse
scattering problem. 

Fig.2 presents the result of the recovery of the potential (\ref{q Ex2sech}).
\begin{figure}
[ptb]
\begin{center}
\includegraphics[
height=3.2941in,
width=4.3846in
]%
{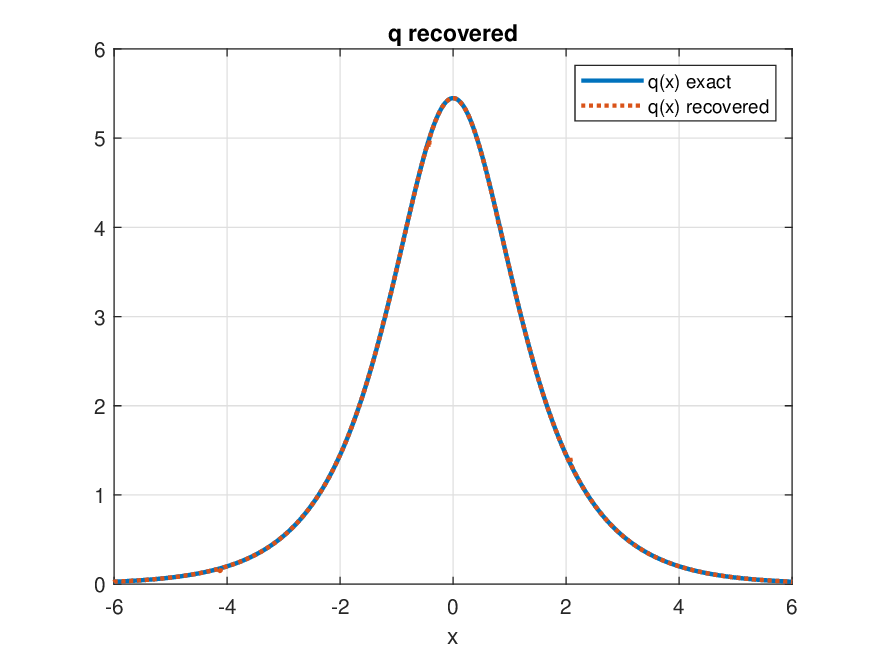}%
\caption{Potential from Example 2 recovered with the maximum absolute error
$0.25$.}%
\label{Fig2}%
\end{center}
\end{figure}
$N$ in (\ref{X}) resulted in $N=65$. Five eigenvalues added 20 equations,
obtained from (\ref{s3}) and (\ref{s4}). Thus, matrix $A$ in (\ref{main sys})
resulted of the dimension $16020\times264$. The maximum absolute error of the
recovered potential was $0.25$. Fig.3 shows the real and imaginary parts of
the computed coeffcient $b_{0}(x)$. They were used for recovering $q(x)$ by
the formula (\ref{q from b0}).%

\begin{figure}
[ptb]
\begin{center}
\includegraphics[
height=3.2629in,
width=4.3431in
]%
{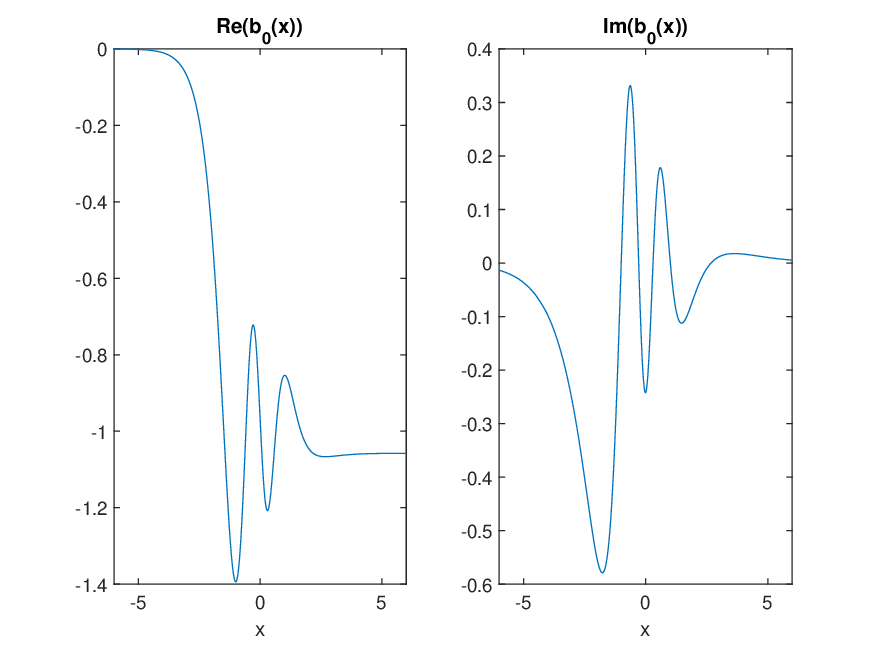}%
\caption{Real and imaginary parts computed of the coefficient $b_{0}(x)$ from
Example 2.}%
\label{Fig3}%
\end{center}
\end{figure}

\textbf{Example 3. }Consider the potential
\begin{equation}
q(x)=\frac{\mu}{\left(  \cosh x\right)  ^{\frac{\pi}{3}}}-e^{-\left(
x-2\right)  ^{2}},\quad\mu=\frac{\pi}{7}.\label{q Example 2}%
\end{equation}
The scattering coefficients $\mathbf{a}(\rho)$ and$\,\mathbf{b}(\rho)$ were
computed following Section \ref{SectDirect} at $4000$ points $\rho\in\left[
-30,30\right]  $. The value of the parameter $N$ (defining the number of terms
in the considered partial sums of the SPPS), chosen according to Remark
\ref{Rem choice of N}, resulted in $N=171$. Fig. 4 depicts the expression
$\left\vert \mathbf{a}(\rho)\right\vert ^{2}+\left\vert \mathbf{b}%
(\rho)\right\vert ^{2}-1$ with the computed scattering coefficients for
$\rho\in\left[  -30,30\right]  $. The exact value of this expression is zero
(relation (\ref{sum of modules squared =1})), while the result of the
computation is of order $10^{-8}$, that gives an idea of the accuracy of the
computed $\mathbf{a}(\rho)$ and$\,\mathbf{b}(\rho)$.%

\begin{figure}
[ptb]
\begin{center}
\includegraphics[
height=3.2223in,
width=4.2947in
]%
{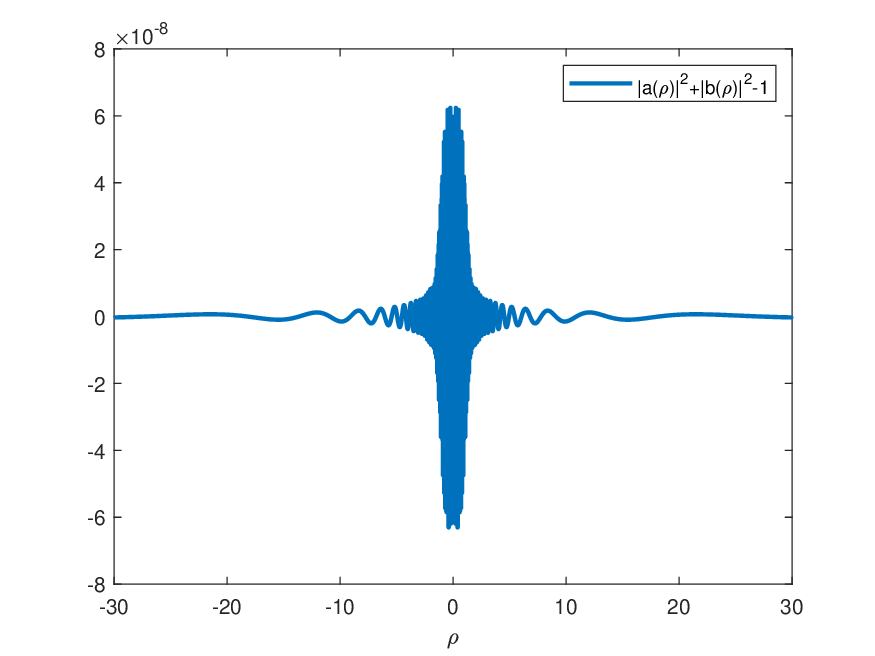}%
\caption{Expression $\left\vert \mathbf{a}(\rho)\right\vert ^{2}+\left\vert
\mathbf{b}(\rho)\right\vert ^{2}-1$ with computed scattering coefficients
$\mathbf{a}(\rho)$ and$\,\mathbf{b}(\rho)$, evaluated at $\rho\in\left[
-30,30\right]  $. }%
\label{Fig4}%
\end{center}
\end{figure}

Two eigenvalues were detected
\[
\rho_{1,2}=\pm0.424731737929926+0.0340968987198153i.
\]

Now we use the computed scattering data for solving the inverse scattering
problem. Fig. 5 presents the result of the recovery of the potential
(\ref{q Example 2}). $N$ in (\ref{X}), which defines the number of the unknown
coefficients in each SPPS representation, resulted in $N=64$, while the
scattering coefficients $\mathbf{a}(\rho)$ and$\,\mathbf{b}(\rho)$ for
(\ref{s1}) and (\ref{s2}) were computed at $4000$ points $\rho\in\left[
-30,30\right]  $. Two eigenvalues added eight equations, obtained from
(\ref{s3}) and (\ref{s4}). Thus, matrix $A$ in (\ref{main sys}) resulted of
the dimension $16008\times260$, its condition number being approximately
$29212$. The maximum absolute error of the recovered potential was
$9.1\cdot10^{-4}$.%

\begin{figure}
[ptb]
\begin{center}
\includegraphics[
height=4.3794in,
width=5.8358in
]%
{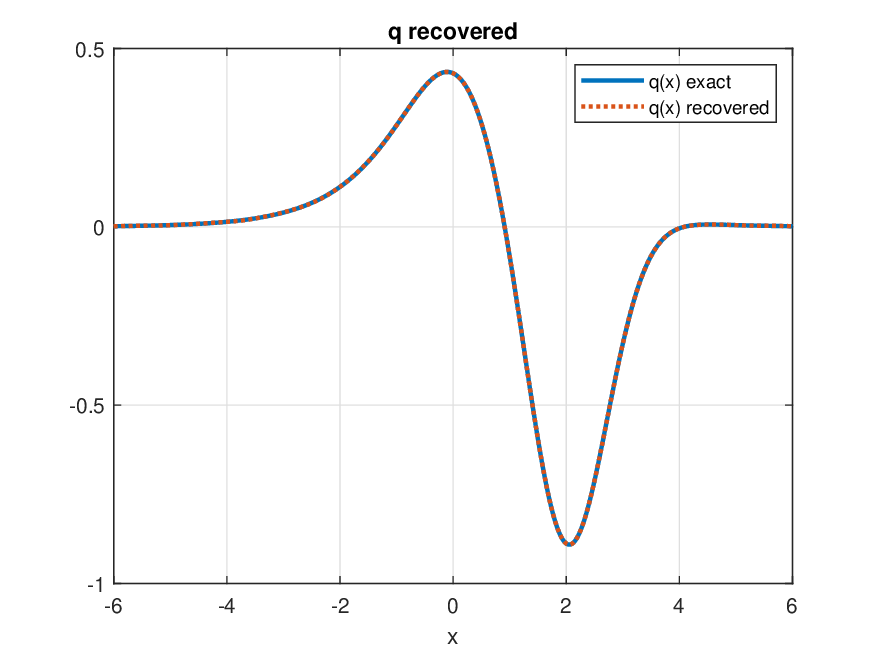}%
\caption{Potential from Example 3, recovered with the maximum absolute error
$9.1\cdot10^{-4}$.}%
\label{Fig5}%
\end{center}
\end{figure}

\textbf{Example 4. }Consider the potential \cite{Civelli et al 2015}, \cite{Le
et al 2014}, \cite{Xiao et al 2002}
\begin{equation}
q(x)=-\frac{4\sqrt{2}(\sqrt{2}-1)}{(\sqrt{2}-1)^{2}e^{-2\sqrt{2}x}%
+e^{2\sqrt{2}x}}.\label{q Ex 4}%
\end{equation}
When solving the direct problem, the computation of the truncation parameter
$N$, performed following Remark \ref{Rem choice of N}, delivered $N=47$. With
this choice of $N$, the unique eigenvalue was detected $\rho_{1}%
=1.414213562373094i$, which coincides with $i\sqrt{2}$ up to $1.5\cdot
10^{-15}$. Again, the location of the roots of the polynomial of the variable
$z$ (see Remark \ref{Rem eigenvalues}) was performed with the aid of the
Matlab routine `roots'. Here we mention that in \cite{Civelli et al 2015},
\cite{Le et al 2014}, \cite{Xiao et al 2002}, in fact, the potential $q(x)$
was extended onto the negative half-axis by zero. The exact reflection
coefficient which was used in those papers corresponds to such a truncated
potential, as can be deduced from the original work \cite{Song et al 1983},
where this example was developed. In that case, unlike the case considered
here, there is no eigenvalue. 

The scattering coefficients $\mathbf{a}(\rho)$ and$\,\mathbf{b}(\rho)$ were
computed following Section \ref{SectDirect} at $4000$ points $\rho\in\left[
-30,30\right]  $. The computed scattering data were used as the input data for
the inverse scattering problem. Fig. 6 presents the result of the recovery of
the potential (\ref{q Ex 4}). The maximum absolute error of the recovered
potential resulted in $9.5\cdot10^{-7}$. %

\begin{figure}
[ptb]
\begin{center}
\includegraphics[
height=3.9323in,
width=5.2338in
]%
{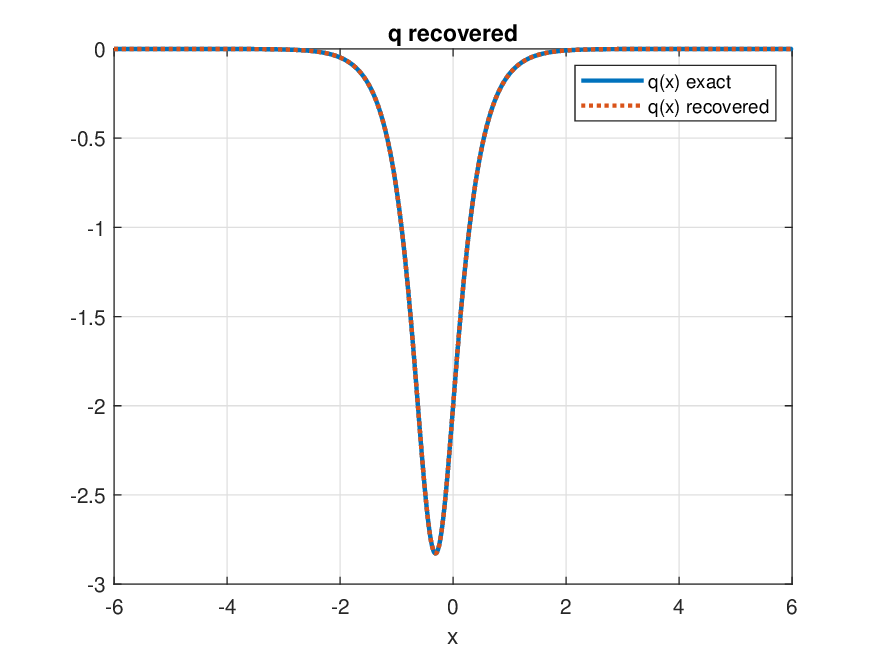}%
\caption{Potential from Example 4, recovered with the maximum absolute error
$9.5\cdot10^{-7}$.}%
\label{Fig6}%
\end{center}
\end{figure}

\section{Conclusions\label{Sect Conclusions}}

The Zakharov-Shabat system with a real-valued decaying potential is
considered. For its Jost solutions power series representations are obtained
in terms of a mapped spectral parameter. The solution of the direct scattering
problem thus reduces to the computation of the coefficients of the series and
location of zeros of a resulting polynomial inside of the unit disk. In its
turn the inverse scattering problem reduces to the solution of a system of
linear algebraic equations, and the potential is recovered from the first
component of the solution vector. The proposed method for solving both the
direct and inverse problem leads to easily implemented, direct and accurate
algorithms. 

\textbf{Funding information }CONAHCYT, Mexico, grant \textquotedblleft Ciencia
de Frontera\textquotedblright\ FORDECYT - PRONACES/ 61517/ 2020.

\textbf{Data availability} The data that support the findings of this study
are available upon reasonable request.

\textbf{Conflict of interest }This work does not have any conflict of interest.

\end{document}